# AI-assisted Learning for Electronic Engineering Courses in High Education

**Thanh Nguyen Ngoc[1], Quang Nhat Tran[2], Arthur Tang[3], Bao Nguyen[4], Thuy Nguyen[5], Thanh Pham[6]**

**Abstract**: This study evaluates the efficacy of ChatGPT as an AI teaching and learning support tool in an integrated circuit systems course at a higher education institution in an Asian country. Various question types were completed, and ChatGPT responses were assessed to gain valuable insights for further investigation. The objective is to assess ChatGPT's ability to provide insights, personalized support, and interactive learning experiences in engineering education. The study includes the evaluation and reflection of different stakeholders: students, lecturers, and engineers. The findings of this study shed light on the benefits and limitations of ChatGPT as an AI tool, paving the way for innovative learning approaches in technical disciplines. Furthermore, the study contributes to our understanding of how digital transformation is likely to unfold in the education sector.

**Additional Keywords and Phrases:** ChatGPT, Generative AI, Digital transformation, engineering education, tutorial design, peer-assisted learning, AI-assisted learning, integrated circuit education.

[1] School of Science, Engineering, and Technology, RMIT University Vietnam
e-mail: thanh.nguyenngoc@rmit.edu.vn

[2] School of Science, Engineering, and Technology, RMIT University Vietnam
e-mail: quang.tran26@rmit.edu.vn

[3] School of Science, Engineering, and Technology, RMIT University Vietnam
e-mail: arthur.tang@rmit.edu.vn

[4] School of Science, Engineering, and Technology, RMIT University Vietnam
e-mail: bao.nguyenthien@rmit.edu.vn

[5] School of Science, Engineering, and Technology, RMIT University Vietnam
e-mail: thuy.nguyen43@rmit.edu.vn

[6] School of Science, Engineering, and Technology, RMIT University Vietnam
e-mail: thanh.pham@rmit.edu.vn

## 1 BACKGROUND

There is a growing interest in using artificial intelligence (AI) to improve teaching and learning [1, 2]. Generative AI tools like ChatGPT understand and generate human-like responses in real-time [3]. This enables dynamic and interactive learning environments where students can have conversations, ask questions, and get instant feedback from AI virtual assistants [4]. These tools have great potential in offering personalized guidance, aiding problem-solving, and providing extra resources for individual learning needs [5].

However, the specific path for implementing these changes remains unclear to many researchers [6]. For instance, their application in programming and engineering education has been relatively limited. Investigating generative AI tools like ChatGPT in these fields is important due to their unique challenges. Programming and engineering require precise understanding and tailored feedback [7, 8]. Evaluating ChatGPT's performance helps identify limitations and develop specialized AI models, ensuring effective AI support in these subjects.

This paper explores the transformation of learning and teaching in engineering courses and examines the impact of AI tools on the education sector. Specifically, we attempt to evaluate ChatGPT as an AI teaching and learning support tool in an introductory integrated circuit systems course at a higher education institution in an Asia country. We aim to assess ChatGPT's ability to provide insights, personalized support, and interactive learning experiences. It could shed light on the benefits and limitations of ChatGPT as an AI tool in engineering education, paving the way for innovative learning approaches in technical disciplines [9]. We also examine the usefulness of ChatGPT from different perspectives such as from students, lecturers, and engineers.

By examining the current situation, discussing benefits and challenges, and providing examples, this study shows how AI-assisted learning can transform engineering education and encourages educators to adopt innovative approaches for training future engineers [10].

## 2 LITERATURE REVIEW

Personalized learning with AI has emerged as a powerful tool in education, revolutionizing the way students engage with content and acquire knowledge [11, 12]. AI-powered systems can analyze vast amounts of data, including student performance, interests, and learning patterns, to provide tailored recommendations and learning paths [13]. This personalized approach enables students to learn at their own pace, focus on areas of interest, and receive targeted support and feedback [12]. One of the key advantages of personalized learning experiences with AI is the ability to provide adaptive instruction [12]. AI algorithms can dynamically adjust the difficulty level and content delivery based on the student's progress and mastery of concepts [12]. This ensures that learners are appropriately challenged and engaged, maximizing their learning outcomes.

Furthermore, AI can facilitate interactive and engaging learning experiences through virtual assistants, chatbots, and intelligent tutoring systems [13]. Such tools can engage in real-time conversations, answer student questions, provide explanations, and offer additional resources. The instant feedback and guidance from AI systems foster active learning and critical thinking [11].

However, it is essential to address the challenges associated with personalized learning experiences with AI [6]. Privacy and data security concerns must be carefully addressed to ensure the ethical use and protection of student data. Therefore, to fully leverage the power of LLM, educators need training and support to effectively integrate AI technologies into the learning environment and to understand how to interpret and use AI-generated insights [4]. This paper addresses the literature gap by offering principles and suggestions that can guide educators in moving forward in that direction.

## 3 RESEARCH METHODOLOGY

### 3.1 Case Study

An engineering course in the Bachelor of Engineering program at a university in an Asian country. Despite ChatGPT is not officially available in the country, its usage has been in the public eye ranging from governments, university students, academics, and engineers [14-16]. The course is an introduction to semiconductor and integrated circuit (IC) design, a specialized course for third- and fourth-year students. The course involves theoretical and coding works. Coding questions require definite solutions that can be validated through graphical waveforms or deployed on an electronic hardware board. The study was done one month after the actual course finished. This study serves as an evaluation rather than a reflection of actual course delivery.

There are 24 questions and exercises selected for the study. The full list of questions is found in Table 4 of the Appendix 1. These were categorized into various levels, as indicated in Table 1. They ranged from straightforward tasks to more complex ones with graphical and hardware verification required. For any original questions with either description or requirements involving graphical illustrations, they were completely rewritten before being inputted into ChatGPT as the tool only supports text-based prompts.

Finally, we incorporated a selection of assessment questions and exercises in this study (Category 3). The assessment questions presented in this paper were modified to conceal the original requirements while maintaining the same difficulty level. The outcome of this selection was to check if existing assessments needed to be redesigned to ensure they were ChatGPT-proof.

Table 1: Categories of questions and exercises

| # | Category | Description |
|---|----------|-------------|
| 1 | Theory questions: definition, categorization, critical analysis, and calculation | These questions are designed to assess students' understanding of specific definitions. Students are evaluated based on their ability to demonstrate, explain, and critically analyze the provided information. |
| 2 | Programming questions/exercises: code snippets for a requirement, or a full code for complex tasks or specifications. | These questions require students to write either a small code snippet or a complete code (100-200 lines of code) in the System Verilog programming language to fulfil a given requirement. To verify the accuracy of the provided code, we need some graphical waveforms (produced by another software), and in some cases can be verified using an electric board. |
| 3 | Assessment programming exercise | The exercises in this category are more sophisticated exercises that require a bit longer solution and combine knowledge and skills from various weeks instead of one as in Category 2. These exercises must come together with a hardware verification. |

## 3.2 Multiple perspectives on the use of AI for learning.

The study examines the usage of ChatGPT from multiple perspectives of different participants. The inclusion of different points of view can help to evaluate how we can optimize the usage of ChatGPT for educational contexts. Participants are:

- Student #1: a student who completed this course a month ago. He was aware of ChatGPT but had not used the tool in any context before this study. He can assess the correctness of some questions.
- Student #2: a student who completed this course and the follow-up course a year ago. He is currently doing a capstone project and interning at an IC design company. He did use ChatGPT for some exploration but not in the context of this course before this study. His technical skills were proven to be stronger than those of student #1.

Both students play the role of learners, and each will try 12 questions.

- Lecturer #3: a lecturer with 10 years of teaching experience in this topic at the university. He used ChatGPT in some limited capacity but did not try ChatGPT for this course until this study.
- Engineer #4: An engineer with 5 years of working experience in IC design. He is currently working for a global company based in the country. He was aware of ChatGPT but did not use it before this study. He participated in training new interns at his company and hence can provide insights into the applicability of ChatGPT in training.

Both the Engineer and the Lecturer play the role of learners, as well as commenting on the questions/prompts from the two students earlier. This is to evaluate the quality of the prompts generated by different learners.

## 3.3 AI-assisted learning flow

In this study, an AI-assisted learning flow is used by all participants as seen in Figure 1. Learners first will contextualize the question, provide an initial prompt to ChatGPT and collect the generated response. In this flow, we use three levels of validation for answers. They include a) checking that the answers are in the context of the question, b) examining the code generated manually, and c) finally verifying the accuracy of the code by producing some graphical waveforms or deploying it on an electronic board used in this course. For each level, if the answers were not satisfactory, learners rephrased questions, added more details and descriptions, provided contexts, and even needed to challenge ChatGPT until an acceptable answer was received in the next iteration. For theory questions, learners can validate by comparing the answers from ChatGPT with existing solutions from the lecturer. It is critical to mention that between iterations or attempts, learners might go through the process of manually analyzing and troubleshooting the issues or errors from the answer or the code. This step is important as learners can achieve the learning outcomes by going through the practices themselves.

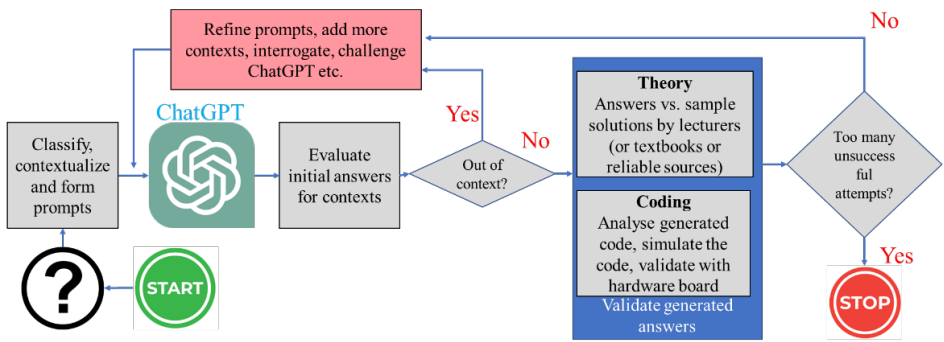

Figure 1- AI-assisted learning flow using ChatGPT.

Figure 2 shows an example response to a question from Category 2, where learners consult ChatGPT on a coding question. The answer includes a code window with a Copy button allowing users to transfer to their target Integrated Development Environment (IDE). Additionally, an accompanying short code explanation can be used to explain the construction of the code. Such an explanation is a significant tool for instructing users about code construction. For verification, learners need to verify the design by producing a waveform and then comparing it with the specification.

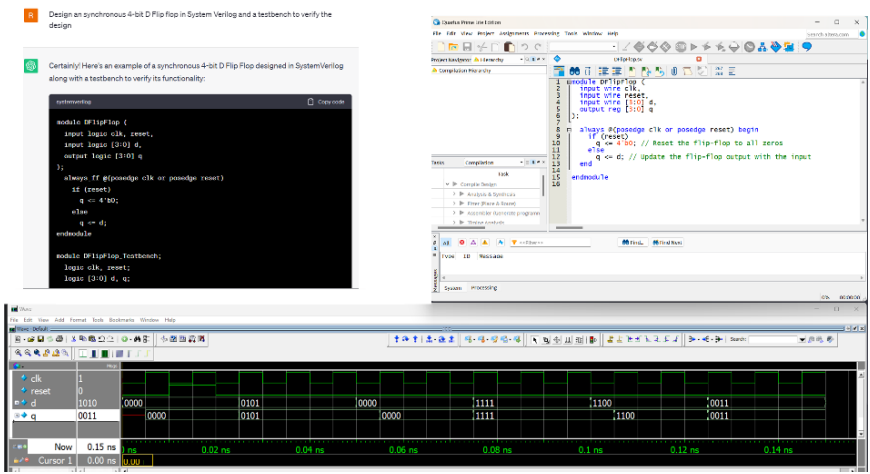

Figure 2 – Sample code generated from ChatGPT (top left) is transferred to the IDE (top right) and verified via graphical waveforms (bottom) generated by the IDE.

## 4  FINDING AND DISCUSSION

### 4.1  Evaluating the answers provided by ChatGPT

To produce a consistent evaluation, this study defined and adopted the rating system described in Table 2 for all participants to evaluate the generated response from ChatGPT. Each participant will use the rating to evaluate the answers for the first two iterations.

Table 2: Definition of evaluation rating.

| Rating | Description |
|---|---|
| I (invalid) | The answer is not relevant or out of context, which needs rephrase, recontextualize and redone. |
| A (applicable) | The answer is clear and correct. |
| C (close) | The answer is clear and close but changes in prompts are still required (via rephrasing, more contexts or challenging ChatGPT). |
| M (misleading) | The answer is clear, but the code contains misleading information (e.g., the value of variables is unrealistic). Users must provide contexts, and challenge ChatGPT to correct the misleading info in the next iteration. |
| F (functional) | The answer is clear, the final code is functional on the designated hardware and meets the requirements. |
| U (unsuccessful) | The answer cannot be reached after so many attempts. |

The evaluation for all questions done by two students is summarized in Table 3. Their chat logs can be found in Appendix 2.

Table 3: Self-evaluation by two students on the answers from ChatGPT for their questions using the rating definition from Table 2.

| Question # | Category | Student | Initial Rate | No of Attempt | Final Rate |
|---|---|---|---|---|---|
| 1 | 1 | 1 | A | 1 | A |
| 2 | 1 | 1 | A | 1 | A |
| 3 | 1 | 1 | A | 1 | A |
| 4 | 1 | 1 | A | 1 | A |
| 5 | 3 | 1 | A | 1 | A |
| 6 | 1 | 1 | A | 1 | A |
| 7 | 1 | 1 | M | 1 | M |
| 8 | 2 | 1 | I | N/A | I |
| 9 | 2 | 1 | F | 1 | F |
| 10 | 2 | 1 | I | N/A | I |
| 11 | 2 | 1 | M | N/A | M |
| 12 | 3 | 1 | C | 1 | C |
| 13 | 1 | 2 | I | 1 | I |
| 14 | 1 | 2 | A | 1 | A |
| 15 | 1 | 2 | C | 1 | C |
| 16 | 2 | 2 | A | 1 | A |
| 17 | 1 | 2 | A | 1 | A |
| 18 | 1 | 2 | A | 1 | A |
| 19 | 1 | 2 | M | N/A | M |
| 20 | 2 | 2 | C | 1 | C |
| 21 | 2 | 2 | A | 1 | A |
| 22 | 2 | 2 | C | 14 | F |
| 23 | 3 | 2 | C | 12 | U |
| 24 | 3 | 2 | C | 1 | C |

First, we can discuss the way the prompts are formed:

- Most of the prompts are within the context of the course, except for two invalid prompts (#8 and #10). This could be explained due to the fact the prompts are carefully designed by the students, who have completed the actual course. This however can be an issue for new learners in their first trials. The two invalid prompts were outside of the scope of this course. After being reviewed with the Lecturer and the Engineer, the two are ignored.
- From the perspective of the Lecturer and the Engineer, most of the initial prompts by the two students lack details and failed to boost the likelihood of getting a valid answer. Especially for coding questions, the current prompts are too short, students must provide a longer detail for each technical component in the prompt. Students can consider writing their prompt in a longer format with multiple paragraphs. For example, they did not specify the programming language System Verilog in some prompts. Or they did not describe the type of signals for the intended modules, which makes ChatGPT guess and hence produce a misleading result.
- The ability to evaluate the response is extremely critical. Even though both students completed the course already, student #1 could not show his ability to extract the right answers from the response of ChatGPT. In many cases, he just took whatever ChatGPT generated without removing redundant information. This raises the concern that the new learners might struggle in evaluating the answers from ChatGPT if they use it to assist their learning. Student #2 on the other hand with a stronger technical background, was able to follow up with his questions and narrow down good answers.

Regarding the quality of the answers themselves, we also observe and discuss as follows:
- ChatGPT can produce good answers for theory, fact-based questions (#1 to #6, #14, #16 to #18, and #21). The responses include useful information and details and allow students to follow up if required. It is expected that ChatGPT was not able to answer question #13 using the current information as the model was trained with the data up to the year 2021.
- ChatGPT is not capable of producing good responses for questions that require graphical illustrations answers (#7 and #19). These are commonly found questions in this course but have proven to be not ChatGPT compatible yet.
- For simple coding exercises either from tutorials or assessments, ChatGPT managed to produce a close answer and sometimes functional code (#9 and #20)
- For sophisticated coding exercises, ChatGPT produced misleading answers if the prompt is too brief (#11) but managed to generate a better code with multiple prompt refinements (#22). It is important to note the students also analyzed and improve the generated code manually during the refinement. This is an essential part of the flow in Table 1.
- For assessment-level exercises (#23 and #24), responses are a mixed success. Despite many attempts, the final code is either just closed or unsuccessful used. Manual troubleshoots or even re-designed are required for these types of questions.
- All participants agreed on the fact that ChatGPT can be used for obtaining new knowledge in the chosen topic, but there are many concerns that need to be addressed in future works. Long reflections of each can be found in the Appendix 3.

## 4.2 Discussion

This study demonstrates the ability of ChatGPT as an AI teaching and learning support tool in an introductory integrated circuit systems course while analyzing its capabilities and limitations. These findings contribute to the ongoing debate surrounding the integration of AI technologies in the education [17, 18].

The results indicate that ChatGPT is proficient in generating accurate and informative responses for theory and fact-based questions, enhancing students' understanding of core concepts [2, 19]. However, it struggles with questions that require graphical illustrations, a common requirement in technical courses. This highlights the need for further development and improvement in AI models to effectively handle visual content.

In terms of coding exercises, ChatGPT demonstrates promising performance for more straightforward tasks or designed modules, providing close answers, and functional and verifiable code. However, its ability to handle complex coding exercises varies. ChatGPT may produce misleading answers if the prompt is brief, but with multiple prompt refinements, it shows potential for generating better code. Nevertheless, human intervention and students' critical thinking skills remain crucial in refining and improving the generated code [2, 19]. The students need to go through manual troubleshooting and refinement, which can be decisive for them to achieve the learning outcomes intended.

Future research directions should focus on addressing the limitations identified in this study. This includes improving the ability of AI models to handle graphical queries, refining the performance of AI in complex coding exercises and assessment-level tasks, and exploring ways to enhance the collaboration between AI tools and human instructors [5, 19].

Furthermore, future studies could investigate the impact of AI integration on learning outcomes and student engagement [4, 20]. Understanding how AI tools like ChatGPT influence students' learning experiences, knowledge retention, and problem-solving skills would provide valuable insights for optimizing their use in the education [17].

Finally, the proposed AI-assisted learning flow in this study can be further studied for other courses. Additionally, the benefits of having experts (The Lecturer and the Engineer) providing comments on the students' ChatGPT usage can be generalized into another type of support for students learning. Lecturers can renovate the course structure to incorporate additional activities focused on training students to effectively use ChatGPT. This approach nurtures their ability to harness the power of ChatGPT for assistance, promoting a deeper understanding of the subject matter and enhancing their overall learning outcomes.

## 5 CONCLUSION

In conclusion, the evaluation of ChatGPT as an AI teaching and learning support tool in the integrated circuit systems course offers valuable insights into the potential of AI technologies to enhance education. The findings highlight both the capabilities and limitations of ChatGPT, providing a foundation for further exploration and improvement.

By recognizing the identified limitations, educators can take proactive steps to address them. For instance, efforts can be made to develop AI models that are better equipped to handle graphical queries, enabling students to receive comprehensive and accurate responses that incorporate visual elements. Additionally, focusing on refining the performance of AI in complex coding exercises and assessment-level tasks can lead to more reliable and satisfactory outcomes. This involves advancing AI models to understand the intricacies of programming and engineering, allowing for precise and contextually appropriate code generation.

*Education Research and Practice,* Review vol. 24, no. 2, pp. 392-393, 2023, doi: 10.1039/d3rp90003g.

[3] K. Dwivedi *et al.*, ""So what if ChatGPT wrote it?" Multidisciplinary perspectives on opportunities, challenges and implications of generative conversational AI for research, practice and policy," *International Journal of Information Management,* vol. 71, p. 102642, 2023.

[4] M. E. Emenike and B. U. Emenike, "Was This Title Generated by ChatGPT? Considerations for Artificial Intelligence Text-Generation Software Programs for Chemists and Chemistry Educators," *Journal of Chemical Education,* Review vol. 100, no. 4, pp. 1413-1418, 2023, doi: 10.1021/acs.jchemed.3c00063.

[5] C. Kooli, "Chatbots in Education and Research: A Critical Examination of Ethical Implications and Solutions," *Sustainability (Switzerland),* Article vol. 15, no. 7, 2023, Art no. 5614, doi: 10.3390/su15075614.

[6] J. Lodge, K. Thompson, and L. Corrin, "Mapping out a research agenda for generative artificial intelligence in tertiary education," 2023.

[7] D. Adebanjo, T. Laosirihongthong, P. Samaranayake, and P. L. Teh, "Key Enablers of Industry 4.0 Development at Firm Level: Findings from an Emerging Economy," *IEEE Transactions on Engineering Management,* Article vol. 70, no. 2, pp. 400-416, 2023, doi: 10.1109/TEM.2020.3046764.

[8] E. T. Santos and S. L. Ferreira. *An Introductory BIM Course for Engineering Students, Lecture Notes on Data Engineering and Communications Technologies*, vol. 146, pp. 880-890, 2023.

[9] M. C. Rillig, M. Ågerstrand, M. Bi, K. A. Gould, and U. Sauerland, "Risks and Benefits of Large Language Models for the Environment," *Environmental Science and Technology,* Short Survey vol. 57, no. 9, pp. 3464-3466, 2023, doi: 10.1021/acs.est.3c01106.

[10] G. Cooper, "Examining Science Education in ChatGPT: An Exploratory Study of Generative Artificial Intelligence," *Journal of Science Education and Technology,* Article vol. 32, no. 3, pp. 444-452, 2023, doi: 10.1007/s10956-023-10039-y.

[11] M. M. Rahman and Y. Watanobe, "ChatGPT for Education and Research: Opportunities, Threats, and Strategies," *Applied Sciences (Switzerland),* Article vol. 13, no. 9, 2023, Art no. 5783, doi: 10.3390/app13095783.

[12] C. L. Lai, "Exploring University Students' Preferences for AI-Assisted Learning Environment: A Drawing Analysis with Activity Theory Framework," *Educational Technology and Society,* Article vol. 24, no. 4, pp. 1-15, 2021. [Online]. Available: https://www.scopus.com/inward/record.uri?eid=2-s2.0-85117959755&partnerID=40&md5=22a975a341e7cc3f816be30e6527abc4.

[13] C. T. Cheng *et al.*, "Artificial intelligence-based education assists medical students' interpretation of hip fracture," *Insights into Imaging,* Article vol. 11, no. 1, 2020, Art no. 119, doi: 10.1186/s13244-020-00932-0.

[14] SGGP. "ChatGPT should be effectively exploited in education." https://en.sggp.org.vn/chatgpt-should-be-effectively-exploited-in-education-post100239.html (accessed 4 July 2023, 2023).

[15] D. Linh. "VinUni to incorporate ChatGPT into educational process." https://vneconomy.vn/vinuni-to-incorporate-chatgpt-into-educational-process.htm (accessed 4 July 2023, 2023).

[16] X.-Q. Dao and N.-B. Le, "Investigating the Effectiveness of ChatGPT in Mathematical Reasoning and Problem Solving: Evidence from the Vietnamese National High School Graduation Examination," *arXiv preprint arXiv:2306.06331,* 2023.

[17] [S. Fergus, M. Botha, and M. Ostovar, "Evaluating Academic Answers Generated Using ChatGPT," *Journal of Chemical Education,* Article vol. 100, no. 4, pp. 1672-1675, 2023, doi: 10.1021/acs.jchemed.3c00087.

[18] C. K. Lo, "What Is the Impact of ChatGPT on Education? A Rapid Review of the Literature," *Education Sciences,* Review vol. 13, no. 4, 2023, Art no. 410, doi: 10.3390/educsci13040410.

[19] V. Benuyenah, "Commentary: ChatGPT use in higher education assessment: Prospects and epistemic threats," *Journal of Research in Innovative Teaching and Learning,* Note vol. 16, no. 1, pp. 134-135, 2023, doi: 10.1108/JRIT-03-2023-097.

[20] B. Gregorcic and A. M. Pendrill, "ChatGPT and the frustrated Socrates," *Physics Education,* Article vol. 58, no. 3, 2023, Art no. 035021, doi: 10.1088/1361-6552/acc299.
9

# APPENDICES

## Appendix 1 – Full list of questions

Table 4 – List of questions/exercises from Student 1 and Student 2

| # | Student | Question | Category |
|---|---------|----------|----------|
| 1 | 1 | What is a transistor made of? | 1 |
| 2 | | How is a transistor made? | 1 |
| 3 | | What is a threshold voltage of a transistor? | 1 |
| 4 | | Provide the threshold voltage formula for a MOS transistor | 1 |
| 5 | | Conditions which affect MOSTFET threshold voltage | 3 |
| 6 | | Any obstacle when making smaller FET size | 1 |
| 7 | | Simple AND/OR gate diagram | 1 |
| 8 | | Write an array sort in system Verilog HDL | 2 |
| 9 | | With input clock 50Mhz, write a clock divider to get 10khz clock | 2 |
| 10 | | Write a SystemVerilog HDL debounce function | 2 |
| 11 | | Write SystemVerilog HDL code to send 8 bits with a clock of 20Hz | 2 |
| 12 | | What is First In First Out? Help me write FIFO for array A[2:0] with elements a,b,c | 3 |
| 13 | 2 | What is the latest technology node at present? | 1 |
| 14 | | What is FinFet? How is it compared to older technology? | 1 |
| 15 | | In CMOS, why PMOS is used for Pull-up Network and NMOS is used for Pull-down Network? | 1 |
| 16 | | What is JK flip-flop? How is it different from D flip-flop? How to implement a JK flip-flop in Verilog? | 2 |
| 17 | | What is the significance of standard cell libraries, how standard cell libraries are designed? What are challenges of designing a standard cell library? | 1 |
| 18 | | What is cell characterization in VLSI? | 1 |
| 19 | | How many MUX 2 to 1 needed to make MUX 16 to 1? | 1 |
| 20 | | Design a 4-bit full-adder with input A, B, Cin and output S, Cout using System Verlilog | 2 |
| 21 | | How do I implement an FSM in SystemVerilog? | 2 |
| 22 | | Help me write SystemVerilog code for positive edge detector | 2 |
| 23 | | Help me design a FSM-based pattern detector in SystemVerilog that takes in a 1-bit serial input stream and asserts the output signal when the pattern '1101' is detected. | 3 |
| 24 | | Help me design a stopwatch module that takes a button input for start/stop and another for reset/lap and keep track to 1/10th of a second. When stopwatch is in idle state, pressing start/stop and the stopwatch should start running and count up until 99 seconds and go back to 0. When the stopwatch is running, pressing start/stop button again will pause the stopwatch. Also when the stopwatch is running, pressing lap/reset should store the current time value to a reg. When the stopwatch is paused, pressing start/stop will resume counting and pressing lap/reset will make the stopwatch stop and reset to 0 (idle state). | 3 |

**Appendix 2 – links to prompts and generated answers from ChatGPT**

The chat prompts and results can be accessed via the following links.

Student 1

https://chat.openai.com/share/3d93d4ea-ec55-4a7c-bf22-d2d8da3d953d

Student 2

https://chat.openai.com/share/0275e995-8c5b-47d1-82c2-c0ff16ed6b64

**Appendix 3 – Reflection from participant's post-study**

The following are short excerpts from the participants on how they perceive ChatGPT and their experience using it for answering questions presented in this study.

**Student #1:**

"I have some insights into the strengths and weaknesses of ChatGPT in advance, but it is great to see how well it answers theoretical questions. It provides very detailed answers when the questions are clear, and it mentions specific requirements within the discussed topic. I am amazed by its ability to answer simple coding questions accurately, especially when it explains the code it generates. However, when it comes to complex coding questions, the code generated by ChatGPT can often be misleading or unsuccessful when run. I am considering using it to write submodules for complex questions and then integrating them into my main code. I am still unsure about relying solely on ChatGPT for the entire coding requirements. Overall, I believe the current state of ChatGPT is a useful tool for supporting the study. It is important for users to have some basic knowledge about the field and to double-check the answers before implementing them. I am thinking about using ChatGPT to generate paragraphs for essays or using it as a paraphrasing tool, although I am afraid that is not allowable by the university."

**Student #2:**

"I had higher expectations regarding ChatGPT's ability to answer my theory questions accurately and provide workable code for my exercises. However, the responses from ChatGPT for theories are often outdated and, in many cases, make me skeptical. Frequently, the answers go against common sense, and without a solid background in the topic, it is difficult to determine if the information is incorrect or completely fabricated. Moreover, when I attempt to challenge ChatGPT or prompt it again, it tends to accept what said as correct rather than being assertive, which is disappointing. I also could not get the information source provided by ChatGPT, which is important if I wanna use it for assessment. For coding-related questions, the provided responses are often unclear and hard to follow. While the code may seem legitimate at first glance, most of the time, it cannot be used as is and requires manual modifications. Additionally, when asked to fix coding errors, the subsequent responses might solve one problem but create another, resulting in unusable code after multiple iterations.

Personally, I do not consider ChatGPT to be a very helpful tool, especially in the area of IC design. To detect the wrong information in ChatGPT's response requires one to truly have some understanding of the topic. These people might not even need ChatGPT in the first place. For beginners and learners, ChatGPT can be highly misleading and can undermine someone's foundation on the topic. It however can be used for quick information lookup, but it is generally not trustworthy."

**Lecturer #3**

"I was aware that ChatGPT could provide answers to some tutorial questions from a course of my co-worker but did not try it until this study. The responses from ChatGPT were stunning to me especially its ability to quickly provide the answers in the contexts. Despite the variation in the accuracy of these answers, from my point of view, if learners try to refine their prompts and use critical thinking skills, apply external resources (i.e., books, journals etc.) they can eventually produce an acceptable answer to questions.

I was amazed by the fact that ChatGPT can produce skeleton answers to assessment questions. This fact surely will remind me to revise my assessments to be more ChatGPT prone, or I might need to revisit the assessment framework completely.

Personally, I embrace the fact that ChatGPT will be inevitably used by many people, so I think of it as an opportunity to revise the course structure, content, and delivery to embrace this emerging tool. Training on ChatGPT usage might be an essential part of courses besides their technical content."

**Engineer #4**

"I heard about ChatGPT through the Internet and watched several YouTube videos on how to use it when it was first introduced. I imagined its potential to replace Google search and deliver information in a simpler manner. I also imagined that it could save us a significant amount of time compared to manually searching for information on Google.

After participating in this study, I recognize the strength of ChatGPT as a tool for synthesizing information. It can provide detailed and comprehensive answers to simple questions, enabling users to learn new concepts. However, sometimes the answers are over-explained and can create confusion for people who don't have prerequisite knowledge of the topic.

To apply in the IC design industry, I can imagine new learners can use this tool to search for new information or at least initial answers to the theory. For people who knew the topic up to some level, ChatGPT can be used to provide a summary and concise version. After checking the answers, I suggest learners double-check the information against reliable sources (books, sites, or from experts). For coding questions, I am still not sure how powerful the tool will be but at least we can always confirm if the code is usable as in IC design, we can verify the results via other tools and compare with the specifications. That can be at least an end goal for learners to rely on.

The most concerning thing is the ability of people to check the info generated by ChatGPT. For example, if they do not verify, they might just take a misinformed answer, or something is wrong.

As part of training for new hires or interns, I can see the above usage from ChatGPT. More importantly, we got engineers to supervise these interns so we can provide ways to verify the information generated by ChatGPT. For actual commercial projects at the company, it is prohibited for engineers to use ChatGPT at all, due to the fear that we might leak our confidential information. I hope some in-house customized generative AI to be introduced in the future."